\documentclass[conference]{IEEEtran}
\IEEEoverridecommandlockouts
\usepackage{cite}
\usepackage{amsmath,amssymb,amsfonts}
\usepackage{algorithmic}
\usepackage{graphicx}
\usepackage{textcomp}
\usepackage{xcolor}
\usepackage[utf8]{inputenc}
\usepackage{marvosym}
\usepackage{amsthm}
\usepackage{float}
\usepackage{url}
\usepackage{caption}
\usepackage{enumitem}

\allowdisplaybreaks

\newtheorem{definition}{Definition}
\newtheorem{theorem}{Theorem}
\newtheorem{problem}{Problem}
\newtheorem{lemma}{Lemma}
\newtheorem{remark}{Remark}

\newcommand\recht\operatorname
\usepackage{tikz}
\usetikzlibrary{decorations.pathmorphing}
\def\BibTeX{{\rm B\kern-.05em{\sc i\kern-.025em b}\kern-.08em
    T\kern-.1667em\lower.7ex\hbox{E}\kern-.125emX}}

\title{
\fontsize{14pt}{16pt}\selectfont\textbf{The Privacy Funnel from the Viewpoint of Local Differential Privacy}}

\author{
\IEEEauthorblockN{Milan Lopuhaä-Zwakenberg}
\IEEEauthorblockA{Department of Mathematics and Computer Science\\
Eindhoven University of Technology\\
Eindhoven, the Netherlands\\
email: m.a.lopuhaa@tue.nl}
}

\begin{document}

\IEEEtitleabstractindextext{%
    \begin{abstract}
In the Open Data approach, governments want to share their datasets with the public, for accountability and to support participation. Data must be opened in such a way that individual privacy is safeguarded. The Privacy Funnel is a mathematical approach that produces a sanitised database that does not leak private data beyond a chosen threshold. The downsides to this approach are that it does not give worst-case privacy guarantees, and that finding optimal sanitisation protocols can be computationally prohibitive. We tackle these problems by using differential privacy metrics, and by considering local protocols which operate on one entry at a time. We show that under both the Local Differential Privacy and Local Information Privacy leakage metrics, one can efficiently obtain optimal protocols; however, Local Information Privacy is both  more closely aligned to the privacy requirements of the Privacy Funnel scenario, and more efficiently computable. We also consider the scenario where each user has multiple attributes, for which we define \emph{Side-channel Resistant Local Information Privacy}, and we give efficient methods to find protocols satisfying this criterion while still offering good utility. Exploratory experiments confirm the validity of these methods.
    \end{abstract}
}

\maketitle

\IEEEdisplaynontitleabstractindextext


\section{Introduction}

Under the Open Data paradigm, governments and other public organisations want to share their collected data with the general public. This increases a governments transparency, and it also gives citizens and businesses the means to participate in decision-making, as well as using the data for their own purposes. However, while the released data should be as faithful to the raw data as possible, individual citizen's private data should not be compromised by such data publication.

To state this problem mathematically, let $\mathcal{X}$ be a finite set. Consider a database $\vec{X} = (X_1,\cdots,X_n) \in \mathcal{X}^n$ owned by a data aggregator, containing a data item $X_i \in \mathcal{X}$ for each user $i$ (For typical database settings, each user's data is a vector of attributes $X_i = (X_i^1,\cdots,X_i^m)$; we will consider this in more detail in Section \ref{sec:mult}). This data may not be considered sensitive by itself, however, it might be correlated to a secret $S_i$. 
 The aggregator wants to release the database to the general public while preventing adversaries from retrieving the secret values $S_i$. For instance, $X_i$ might contain the age, sex, weight, skin colour, and average blood pressure of person $i$, while $S_i$ is the presence of some medical condition.
To publicise the data without leaking the $S_i$, the aggregator releases a privatised database $\vec{Y} = (Y_1,\cdots,Y_n)$, obtained from applying a sanitisation mechanism $\mathcal{R}$ to $\vec{X}$. One way to formulate this is by considering the \emph{Privacy Funnel}:

\begin{problem} \label{prob:pf} \emph{(Privacy Funnel, \cite{calmon2017principal})}
Suppose the joint probability distribution of $\vec{S}$ and $\vec{X}$ is known to the aggregator, and let $M \in \mathbb{R}_{\geq 0}$. Then, find the privatization mechanism $\mathcal{R}$ such that $\recht{I}(\vec{X};\vec{Y})$ is maximised while $\recht{I}(\vec{S};\vec{Y}) \leq M$.
\end{problem}

There are two difficulties with this approach:
\begin{enumerate}
\item Finding and implementing good privatization mechanisms that operate on all of $\vec{X}$ can be computationally prohibitive for large $n$, as the complexity is exponential in $n$ \cite{ding2019submodularity}\cite{prasser2014arx}.
\item Taking mutual information as a leakage measure has as a disadvantage that it gives guarantees about the leakage in the average case. If $n$ is large, this still leaves room for the sanitisation protocol to leak undesirably much information about a few unlucky users.
\end{enumerate}

\begin{figure}
\centering
\begin{tikzpicture}[scale = 0.4]
\draw[rounded corners] (1,1) -- (3,1) -- (3,-7) -- (1,-7) -- cycle;
\draw[-] (1,-1)--(3,-1);
\draw[-] (1,-3)--(3,-3);
\draw[-] (1,-5)--(3,-5);
\draw (2,0) node{$S_1$};
\draw (2,-2) node{$S_2$};
\draw (2,-4) node{$\vdots$};
\draw (2,-6) node{$S_n$};
\draw[rounded corners] (6,1) -- (8,1) -- (8,-7) --node[below]{Database} (6,-7) -- cycle;
\draw[-] (6,-1)--(8,-1);
\draw[-] (6,-3)--(8,-3);
\draw[-] (6,-5)--(8,-5);
\draw (7,0) node{$X_1$};
\draw (7,-2) node{$X_2$};
\draw (7,-4) node{$\vdots$};
\draw (7,-6) node{$X_n$};
\draw[rounded corners] (12,1) -- (14,1) -- (14,-7) --node[below, text width = 1.4cm]{Sanitised Database} (12,-7) -- cycle;
\draw[-] (12,-1)--(14,-1);
\draw[-] (12,-3)--(14,-3);
\draw[-] (12,-5)--(14,-5);
\draw (13,0) node{$Y_1$};
\draw (13,-2) node{$Y_2$};
\draw (13,-4) node{$\vdots$};
\draw (13,-6) node{$Y_n$};
\draw[->,decorate,decoration={snake}](8,0) -- node[above]{$\mathcal{Q}$} (12,0);
\draw[->,decorate,decoration={snake}](8,-2) -- node[above]{$\mathcal{Q}$} (12,-2);
\draw[->,decorate,decoration={snake}](8,-6) -- node[above]{$\mathcal{Q}$} (12,-6);
\draw[<->,decorate,decoration={snake}] (3,0) -- (6,0);
\draw[<->,decorate,decoration={snake}] (3,-2) -- (6,-2);
\draw[<->,decorate,decoration={snake}] (3,-6) -- (6,-6);
\draw[rounded corners,dashed] (0,2) --node[above]{Hidden from public}  (9,2) -- (9,-9) -- (0,-9) -- cycle;
\end{tikzpicture}
\captionsetup{font={footnotesize,rm},justification=centering,labelsep=period}
\caption{Model of the Privacy Funnel with local protocols.} 
\label{fig:sit}
\end{figure}
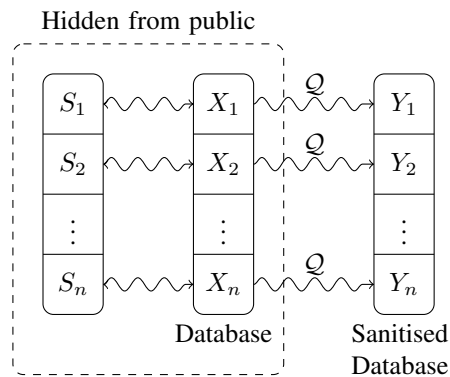

To deal with these two difficulties, we make two changes to the general approach. First, we look at \emph{local} data sanitisation, i.e., we consider optimization protocols $\mathcal{Q}\colon \mathcal{X} \rightarrow \mathcal{Y}$, for some finite set $\mathcal{Y}$, and we apply $\mathcal{Q}$ to each $X_i$ individually; this situation is depicted in Figure \ref{fig:sit}. These can be efficiently implemented. Second, to ensure strong privacy guarantees even in worst-case scenarios, we take stricter notions of privacy, based on Local Differential Privacy (LDP) \cite{kasiviswanathan2011what}. 

The structure of this paper is as follows. In Section \ref{sec:maths}, we define the mathematical setting of our problem. We discuss two privacy notions, LDP and Local Information Privacy (LIP), and discuss their relation to the Privacy Funnel. In Sections \ref{sec:ldp} and \ref{sec:lip}, we show that for a given level of LDP or LIP, respectively, one can efficiently find the optimal sanitisation protocol. In Section \ref{sec:mult}, we consider the setting where every $X_i$ is a vector of attributes, and we show how to make protocols that protect against side-channel attacks. In Section \ref{sec:exp}, we numerically assess the methods presented in this paper.

\section{Mathematical Setting} \label{sec:maths}

The database $\vec{X} = (X_1,\cdots,X_n)$ consists out of a data item $X_i$ for each user $i$, each an element of a given finite set $\mathcal{X}$. Furthermore, each user has sensitive data $S_i \in \mathcal{S}$, which is correlated with $X_i$; again we assume $\mathcal{S}$ to be finite (see Figure \ref{fig:sit}). We assume each $(S_i,X_i)$ is drawn independently from the same distribution $\recht{p}_{S,X}$ on $\mathcal{S} \times \mathcal{X}$ which is known to the aggregator through observing $(\vec{S},\vec{X})$ (if one allows for non-independent $X_i$, then differential privacy is no longer an adequate privacy metric \cite{cuff2016differential}{}\cite{salamatian2020privacy}). The aggregator, who has access to $\vec{X}$, sanitises the database by applying a sanitisation protocol (i.e., a random function) $\mathcal{Q}\colon \mathcal{X} \rightarrow\mathcal{Y}$ to each $X_i$, outputting $\vec{Y} = (Y_1,\cdots,Y_n) = (\mathcal{Q}(X_1),\cdots,\mathcal{Q}(X_n))$. The aggregator's goal is to find a $\mathcal{Q}$ that maximises the information about $X_i$ preserved in $Y_i$ (measured as $\recht{I}(X_i;Y_i)$) while leaking only minimal information about $S_i$.

Without loss of generality we write $\mathcal{X} = \{1,\cdots,a\}$ and $\mathcal{Y} = \{1,\cdots,b\}$ for integers $a,b$. We omit the subscript $i$ from $X_i$, $Y_i$, $S_i$ as no probabilities depend on it, and we write such probabilities as $\recht{p}_x$, $\recht{p}_s$, $\recht{p}_{x|s}$, etc., which form vectors $\recht{p}_{X}$, $\recht{p}_{S|x}$, etc., and matrices $\recht{p}_{X|S}$, etc.

As noted before, instead of looking at the mutual information $\recht{I}(S;Y)$, we consider two different, related measures of sensitive information leakage known from the literature. The first one is an adaptation of LDP, the \emph{de facto} standard in information privacy \cite{kasiviswanathan2011what}:

\begin{definition} \emph{($\varepsilon$-LDP)}
Let $\varepsilon \in \mathbb{R}_{\geq 0}$. We say that $\mathcal{Q}$ satisfies $\varepsilon$-LDP w.r.t. $S$ if for all $y \in \mathcal{Y}$ and all $s,s' \in \mathcal{S}$ one has 
\begin{equation}
\frac{\mathbb{P}(Y = y | S = s)}{\mathbb{P}(Y = y  | S = s')} \leq \textrm{\emph{e}}^{\varepsilon}.
\end{equation}
\end{definition}

This is less strict than the `standard' notion of $\varepsilon$-LDP, which measures the information about $X$ leaked in $Y$. This reflects the fact that we are only interested in hiding sensitive data, rather than all data; it is a specific case of what has been named `pufferfish privacy' \cite{kifer2014pufferfish}. The advantage of LDP compared to mutual information is that it gives privacy guarantees for the worst case, not just the average case. This is desirable in the database setting, as a worst-case metric guarantees the security of the private data of all users, while average-case metrics are only concerned with the average user. Another useful privacy metric is  \emph{Local Information Privacy} (LIP) \cite{jiang2019local}\cite{salamatian2020privacy}, also called Removal Local Differential Privacy \cite{erlingsson2020encode}:

\begin{definition} \emph{($\varepsilon$-LIP)}
Let $\varepsilon \in \mathbb{R}_{\geq 0}$. We say that $\mathcal{Q}$ satisfies $\varepsilon$-LIP w.r.t. $S$ if for all $s \in \mathcal{S}$ and $y \in \mathcal{Y}$ we have 
\begin{equation}
\textrm{\emph{e}}^{-\varepsilon} \leq \frac{\mathbb{P}(Y = y | S = s)}{\mathbb{P}(Y = y)} \leq \textrm{\emph{e}}^{\varepsilon}.
\end{equation}
\end{definition}

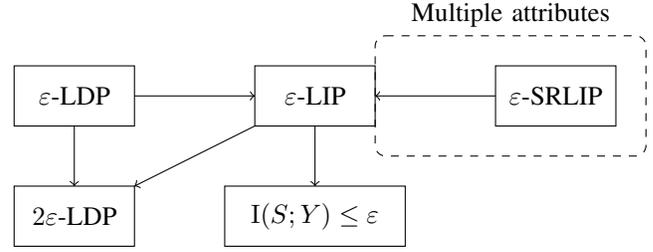
\begin{figure}
\centering
\begin{tikzpicture}[scale = 0.4]
\draw[-] (0,0) -- (4,0) -- (4,-2) -- (0,-2) -- cycle;
\draw (2,-1) node{$\varepsilon$-LDP};
\draw[-] (0,-4) -- (4,-4) -- (4,-6) -- (0,-6) -- cycle;
\draw (2,-5) node{$2\varepsilon$-LDP};
\draw[-] (8,0) -- (12,0) -- (12,-2) -- (8,-2) -- cycle;
\draw (10,-1) node{$\varepsilon$-LIP};
\draw[-] (7,-4) -- (13,-4) -- (13,-6) -- (7,-6) -- cycle;
\draw (10,-5) node{$\recht{I}(S;Y) \leq \varepsilon$};
\draw[-] (16,0) -- (20,0) -- (20,-2) -- (16,-2) -- cycle;
\draw (18,-1) node{$\varepsilon$-SRLIP};
\draw[->] (4,-1) -- (8,-1);
\draw[->] (2,-2) -- (2,-4);
\draw[->] (8,-2) -- (4,-4);
\draw[->] (10,-2) -- (10,-4);
\draw[->] (16,-1) -- (12,-1);
\draw[-,dashed,rounded corners] (12,1) --node[above]{Multiple attributes} (21,1) -- (21,-3) -- (12,-3) -- cycle;
\end{tikzpicture}
\captionsetup{font={footnotesize,rm},justification=centering,labelsep=period}
\caption{Relations between privacy notions. The multiple attributes setting is discussed in Section \ref{sec:mult}.} 
\label{fig:rel}
\end{figure}

Compared to LDP, the disadvantage of LIP is that it depends on the distribution of $S$; this is less relevant in our scenario, as the aggregator, who chooses $\mathcal{Q}$, has access to the distribution of $S$. The advantage of LIP is that is more closely related to an attacker's capabilities: since $\frac{\mathbb{P}(Y = y | S = s)}{\mathbb{P}(Y = y)} = \frac{\mathbb{P}(S = s | Y = y)}{\mathbb{P}(S = s)}$, satisfying $\varepsilon$-LIP means that an attacker's posterior distribution of $S$ given $Y = y$ does not deviate from their prior distribution by more than a factor $\textrm{e}^{\varepsilon}$. The following Lemma outlines the relations between LDP, LIP and mutual information (see Figure \ref{fig:rel}).

\begin{lemma} \emph{(See \cite{salamatian2020privacy})}
Let $\mathcal{Q}$ be a sanitisation protocol, and let $\varepsilon \in \mathbb{R}_{\geq 0}$.
\begin{enumerate}
\item If $\mathcal{Q}$ satisfies $\varepsilon$-LDP, then it satisfies $\varepsilon$-LIP.
\item If $\mathcal{Q}$ satisfies $\varepsilon$-LIP, then it satisfies $2\varepsilon$-LDP, and $\recht{I}(S;Y) \leq \varepsilon$.
\end{enumerate}
\end{lemma}

\begin{remark} \label{rem:prior}
One can choose to employ more stringent privacy metrics for LDP and LIP by demanding that $\mathcal{Q}$ satisfy $\varepsilon$-LIP ($\varepsilon$-LDP) for a set of $\recht{p}_{S,X}$, instead of only one \cite{kifer2014pufferfish}. Letting $\recht{p}_{S,X}$ range over all possible distributions on $\mathcal{S} \times \mathcal{X}$ yields standard LIP (LDP) (i.e., w.r.t. $X$).
\end{remark}

In this notation, instead of Problem \ref{prob:pf} we consider the following problem:

\begin{problem} \label{prob:pf2}
Suppose $\recht{p}_{S,X}$ is known to the aggregator, and let $\varepsilon \in \mathbb{R}_{\geq 0}$. Then, find the sanitisation protocol $\mathcal{Q}$ such that $\recht{I}(X;Y)$ is maximised while $\mathcal{Q}$ satisfies $\varepsilon$-LDP ($\varepsilon$-LIP, respectively) with respect to $S$.
\end{problem}

Note that this problem does not depend on the number of users $n$, and as such this approach will find solutions that are scalable w.r.t. $n$.

\section{Optimizing $\mathcal{Q}$ for $\varepsilon$-LDP} \label{sec:ldp}

Our goal is now to find the optimal $\mathcal{Q}$, i.e., the protocol that maximises $\recht{I}(X;Y)$ while satisfying $\varepsilon$-LDP, for a given $\varepsilon$. We can represent any sanitisation protocol as a matrix $Q \in \mathbb{R}^{b \times a}$, where $Q_{y|x} = \mathbb{P}(Y = y | X = x)$. Then, $Q$ defines a sanitisation protocol $\mathcal{Q}$ satisfying $\varepsilon$-LDP if and only if
\begin{align}
\forall x\colon& \ \sum_y Q_{y|x} = 1,\\
\forall x,y\colon& \ 0 \leq Q_{y|x},\\
\forall s,s',y\colon& \ (Q\recht{p}_{X|s})_y \leq \textrm{e}^{\varepsilon} (Q\recht{p}_{X|s'})_y. \label{eq:ldp}
\end{align}
As such, for a given $\mathcal{Y}$, the set of $\varepsilon$-LDP-satisfying sanitisation protocols can be considered a closed, bounded, convex polytope $\Gamma$ in $\mathbb{R}^{b \times a}$. This fact allows us to efficiently find optimal protocols.

\begin{theorem} \label{thm:ldp}
Let $\varepsilon \in \mathbb{R}_{\geq 0}$. Let $\mathcal{Q}\colon \mathcal{X} \rightarrow \mathcal{Y}$ be the $\varepsilon$-LDP protocol that maximises $\recht{I}(X;Y)$, i.e., the protocol that solves Problem \ref{prob:pf2} w.r.t. LDP.
\begin{enumerate}
\item One has $b \leq a$.
\item Let $\Gamma$ be the polytope described above. Then one can find $\mathcal{Q}$ by maximising a convex function on $\Gamma$.
\end{enumerate}
\end{theorem}

This result is obtained by generalising the results of \cite{kairouz2014extremal}: there this is proven for regular $\varepsilon$-LDP (i.e., w.r.t. $X$), but the arguments given in that proof hold just as well in our situation; the only difference is that their polytope is defined by the $\varepsilon$-LDP conditions w.r.t. $X$, but this has no impact on the proof. Together, these results reduce our problem to a finite optimisation problem: By point 1, we only need to consider $\mathcal{Y} = \mathcal{X}$, and, by point 2, we only need to find the set of vertices of $\Gamma$, a $a(a-1)$-dimensional convex polytope.

One might argue that, since the optimal $\mathcal{Q}$ depends on $\recht{p}_{S,X}$, the publication of $\mathcal{Q}$ might provide an aggregator with information about the distribution of $S$. However, information on the distribution (as opposed to information of individual users' data) is not considered sensitive \cite{lopuhaazwakenberg2019information}. In fact, the reason why the aggregator sanitises the data is because an attacker is assumed to have knowledge about this correlation, and revealing too much information about $X$ would cause the aggregator to use this information to infer information about $S$.

\section{Optimizing $\mathcal{Q}$ for $\varepsilon$-LIP} \label{sec:lip}

If one uses $\varepsilon$-LIP as a privacy metric, one can find the optimal sanitisation protocol in a similar fashion. To do this, we again describe $\mathcal{Q}$ as a matrix, but this time a different one. Let $q \in \mathbb{R}^{b}$ be the probability mass function of $Y$, and let $R \in \mathbb{R}^{a \times b}$ be given by $R_{x|y} = \mathbb{P}(X = x | Y = y)$; we denote its $y$-th row by $R_{X|y} \in \mathbb{R}^a$. Then, a pair $(R,q)$ defines a sanitisation protocol $\mathcal{Q}$ satisfying $\varepsilon$-LIP if and only if
\begin{align}
\forall y \colon& \ 0 \leq q_y,\\
& \ Rq = \recht{p}_X, \label{eq:LIP2}\\
\forall y\colon& \ \sum_x R_{x|y} = 1,\label{eq:LIP3}\\
\forall x,y\colon& \ 0 \leq R_{x|y},\label{eq:LIP4}\\
\forall y,s\colon& \ \textrm{e}^{-\varepsilon}\recht{p}_s \leq \recht{p}_{s|X}R_{X| y} \leq \textrm{e}^{\varepsilon}\recht{p}_s. \label{eq:LIP5}
\end{align}

Note that (\ref{eq:LIP5}) defines the $\varepsilon$-LIP condition, since for a given $s,y$ we have $\frac{\recht{p}_{s|X}R_{X| y}}{\recht{p}_S} = \frac{\mathbb{P}(S=s|Y=y)}{\mathbb{P}(S=s)} = \frac{\mathbb{P}(Y=y|S=s)}{\mathbb{P}(Y=y)}$. (In)equalities (\ref{eq:LIP3}--\ref{eq:LIP5}) can be expressed as saying that for every $y \in \mathcal{Y}$ one has that $R_{X|y} \in \Delta$, where $\Delta$ is the convex closed bounded polytope in $\mathbb{R}^{\mathcal{X}}$ given by
\begin{equation}
\Delta = \left\{v \in \mathbb{R}^{\mathcal{X}}:  \begin{array}{l}
                          \sum_x v_x = 1,\\
                          \forall x: 0 \leq v_x,\\
                          \forall s: \textrm{e}^{-\varepsilon}\recht{p}_s \leq \recht{p}_{s|X}v \leq \textrm{e}^{\varepsilon}\recht{p}_s
                          \end{array}\right\}. \label{eq:delta}
\end{equation}
As in Theorem \ref{thm:ldp}, we can use this polytope to find optimal protocols:

\begin{theorem} \label{thm:lip}
Let $\varepsilon \in \mathbb{R}_{\geq 0}$. Let $\mathcal{Q}\colon \mathcal{X} \rightarrow \mathcal{Y}$ be the $\varepsilon$-LIP protocol that maximises $\recht{I}(X;Y)$, i.e., the protocol that solves Problem \ref{prob:pf2} w.r.t. LIP.
\begin{enumerate}
\item One has $b \leq a$.
\item Let $\Delta$ be the polytope described above, and let $\mathcal{V}$ be its set of vertices. Then one can find $\mathcal{Q}$ by solving a $\#\mathcal{V}$-dimensional linear optimization problem. 
\end{enumerate}
\end{theorem}

This is proven for $\varepsilon = 0$ (i.e., when $S$ and $Y$ are independent) in \cite{rassouli2017perfect}, but the proof works similarly for $\varepsilon > 0$; the main difference is that the equality constraints of their (10) will be replaced by the inequality constraints of our (\ref{eq:LIP5}), but this has no impact on the proof presented there. Since linear optimization problems can be solved fast, again the optimization problem reduces to finding the vertices of a polytope. The advantage of this approach, however, is that $\Delta$ is a $(a-1)$-dimensional polytope, while $\Gamma$ is $a(a-1)$-dimensional. The time complexity of vertex enumeration is linear in the number of vertices \cite{avis1992pivoting}, while the number of vertices can grow exponentially in the dimension of the polyhedron \cite{barany2000polytopes}. Together, this means that the dimension plays a huge role in the time complexity, hence we expect finding the optimum under LIP to be significantly faster than under LDP.

\section{Multiple Attributes} \label{sec:mult}

An often-occuring scenario is that a user's data consists out of multiple attributes, i.e., $X_i = (X^1_i,\cdots,X^m_i) \in \mathcal{X} = \prod_{j=1}^m \mathcal{X}^j$. This can be problematic for our approach for two reasons:
\begin{enumerate}
\item Such a large $\mathcal{X}$ can be problematic, since the computing time for optimisation both under LDP and LIP will depend heavily on $a$.
\item In practice, an attacker might sometimes utilise side channels to access to some subsets of attributes $X_i^j$ for some users. For these users, a sanitisation protocol can leak more information (w.r.t. to the attacker's updated prior information) than its LDP/LIP parameter would suggest.
\end{enumerate}

To see how the second problem might arise in practice, suppose that $X^1_i$ is the height of individual $i$, $X^2_i$ is their weight, and $S_i$ is whether $i$ is obese or not. Since height is only lightly correlated with obesity, taking $Y_i = X_i^1$ would satisfy $\varepsilon$-LIP for some reasonably small $\varepsilon$. However, suppose that an attacker has access to $X^2_i$ via a side channel. While knowing $i$'s weight gives the attacker some, but not perfect knowledge about $i$'s obesity, the combination of the weight from the side channel, and the height from the $Y_i$, allows the attacker to calculate $i$'s BMI, giving much more information about $i$'s obesity. Therefore, the given protocol gives much less privacy in the presence of this side channel.

To solve the second problem, we introduce a more stringent privacy notion called \emph{Side-channel Resistant LIP} (SRLIP), which ensures that no matter which attributes an attacker has access to, the protocol still satisfies $\varepsilon$-LIP with respect to the attacker's new prior distribution. One could similarly introduce SRLDP, and many results will still hold for this privacy measure; nevertheless, since we concluded that LIP is preferable over LDP, we focus on SRLIP. For $J \subset \{1,\cdots,m\}$, we write $\mathcal{X}^J = \prod_{j \in J} \mathcal{X}^j$ and its elements as $x^J$.

\begin{definition} \emph{($\varepsilon$-SRLIP)}. Let $\varepsilon > 0$, and let $\mathcal{X} = \prod_{j=1}^m \mathcal{X}^j$. 
We say that $\mathcal{Q}$ satisfies $\varepsilon$-SRLIP if for every $y \in \mathcal{Y}$, for every $s \in \mathcal{S}$, for every $J \subset \{1,\cdots,m\}$, and for every $x^J \in \mathcal{X}^J$ one has
\begin{equation}
\textrm{\emph{e}}^{-\varepsilon} \leq \frac{\mathbb{P}(Y = y | S = s, X^J = x^J)}{\mathbb{P}(Y = y | X^J = x^J)} \leq \textrm{\emph{e}}^{\varepsilon}.
\end{equation}

\end{definition}

In terms of Remark \ref{rem:prior}, $\mathcal{Q}$ satisfies $\varepsilon$-SRLIP if and only if it satisfies $\varepsilon$-LIP w.r.t. $\recht{p}_{S,X|x^J}$ for all $J$ and $x^J$. Taking $J = \varnothing$ gives us the regular definition of $\varepsilon$-LIP, proving the following Lemma:

\begin{lemma} \label{lem:srlip}
Let $\varepsilon > 0$. If $\mathcal{Q}$ satisfies $\varepsilon$-SRLIP, then $\mathcal{Q}$ satisfies $\varepsilon$-LIP.
\end{lemma}

While SRLIP is stricter than LIP itself, it has the advantage that even when an attacker has access to some data of a user, the sanitisation protocol still does not leak an unwanted amount of information beyond the knowledge the attacker has gained via the side channel. Another advantage is that, contrary to LIP itself, SRLIP satisfies an analogon of the concept of \emph{privacy budget} \cite{dwork2006calibrating}:

\begin{theorem} \label{thm:budget} 
Let $\mathcal{X} = \prod_{j=1}^m \mathcal{X}^j$, and for every $j$, let $\mathcal{Q}^j\colon \mathcal{X}^j \rightarrow \mathcal{Y}^j$ be a sanitisation protocol. Let $\varepsilon^j \in \mathbb{R}_{\geq 0}$ for every $j$. Suppose that for every $j \leq m$, for every $J \subset \{1,\cdots,j-1,j+1,\cdots,m\}$, and every $x^{J} \in \mathcal{X}^{J}$, $\mathcal{Q}^j$ satisfies $\varepsilon^j$-LIP w.r.t. $\recht{p}_{S,X|x^{J}}$. Then $\prod_j \mathcal{Q}^j \colon \mathcal{X} \rightarrow \prod_j \mathcal{Y}^j$ satisfies $\sum_j \varepsilon^j$-SRLIP.
\end{theorem}

The proof is presented in Appendix \ref{app:proof}. This theorem tells us that to find a $\varepsilon$-SRLIP protocol for $\mathcal{X}$, it suffices to find a sanitisation protocol for each $\mathcal{X}^j$ that is $\frac{\varepsilon}{m}$-LIP w.r.t. a number of prior distributions. Unfortunately, the method of finding an optimal $\varepsilon$-LIP protocol w.r.t. one prior $\recht{p}_{S,X}$ of Theorem \ref{thm:lip} does not transfer to the multiple prior setting. This is because this method only finds one $(R,q)$, while by (\ref{eq:LIP2}) we need a different $(R,q)$ for each prior distribution. Therefore, we are forced to adopt an approach similar to the one in Theorem \ref{thm:ldp}. The matrix $Q^j$ (given by $Q^j_{y^j|x^j} = \mathbb{P}(\mathcal{Q}^j(x^j) = y^j$)) corresponding to $\mathcal{Q}^j\colon \mathcal{X}^j \rightarrow \mathcal{Y}^j$ satisfies the criteria of Theorem \ref{thm:budget} if and only if the following criteria are satisfied:

\begin{align}
\forall x^j: & \ \sum_{y^j} Q^j_{y^j|x^j} = 1,\label{eq:rslip1}\\
\forall x ^j,y^j: & \ 0 \leq Q^j_{y^j|x^j},\\
\forall J, x^J, s,y^j: & \ \textrm{e}^{-\varepsilon/m}(Q^j\recht{p}_{X^j|x^{J}})_{y^j} \leq (Q^j\recht{p}_{X^j|s,x^{J}})_{y^j},\\
\forall J, x^J, s,y^j: & \ (Q^j\recht{p}_{X^j|s,x^{J}})_{y^j} \leq \textrm{e}^{\varepsilon/m} (Q^j\recht{p}_{X^j|x^{J}})_{y^j}. \label{eq:rslip5}
\end{align}

Similar to Theorem \ref{thm:ldp}, we can find the optimal $\mathcal{Q}^j$ satisfying these conditions by finding the vertices of the polytope defined by these equations. In terms of time complexity, the comparison to finding the optimal $\varepsilon$-LIP protocol via Theorem \ref{thm:lip} versus finding a $\varepsilon$-SRLIP protocol via Theorem \ref{thm:budget} is not straightforward. The complexity of enumerating the vertices of a polytope is $\mathcal{O}(ndv)$, where $n$ is the number of inequalities, $d$ is the dimension, and $v$ is the number of vertices \cite{avis1992pivoting}. For $\Delta$ of Theorem \ref{thm:lip} we have $d = a-1$ and $n = a+2c$. By contrast, for the polytope defined by (\ref{eq:rslip1}--\ref{eq:rslip5}) satisfies $d = a^j(a^j-1)$ and $n = (a^j)^2+2c\prod_{j' \neq j} (a^{j'}+1)$. Finding $v$ for both these polytopes is difficult, but in general $v \leq \binom{n}{d}$. Since this grows exponentially in $d$, we expect Theorem \ref{thm:budget} to be faster when the $a^j$ are small compared to $a$, i.e., when $m$ is large. We will investigate this experimentally in the next section.

\section{Experiments} \label{sec:exp}\vspace{-0pt}

\begin{figure}
\centering
\includegraphics[width = \linewidth]{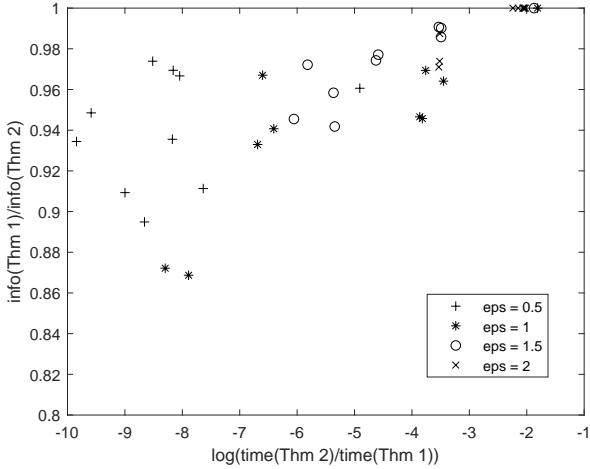}
\captionsetup{font={footnotesize,rm},justification=centering,labelsep=period}
\caption{Comparision of computation time and $\recht{I}(X;Y)$ for $\varepsilon$-LDP protocols found via Theorem \ref{thm:ldp} and $\varepsilon$-LIP protocols found via Theorem \ref{thm:lip}, for random $\recht{p}_{S,X}$ with $c = 2$, $a = 5$, and $\varepsilon \in \{0.5,1,1.5,2\}$.\label{fig:ldplip}}
\end{figure}

We test the feasibility of the different methods and privacy definitions by performing small-scale experiments on synthetic data. All experiments are implemented in Matlab and conducted on a PC with Intel Core i7-7700HQ 2.8GHz and 32GB memory. We compare the computing time for finding optimal $\varepsilon$-LDP and $\varepsilon$-LIP protocols for $c = 2$ and $a = 5$ for 10 random $\recht{p}_{S,X}$, obtained by generating each $\recht{p}_{s,x}$ uniformly from $[0,1]$ and then rescaling. We take $\varepsilon \in \{0.5,1,1.5,2\}$; the results are in Figure \ref{fig:ldplip}. As one can see, Theorem \ref{thm:lip} gives significantly faster results than Theorem \ref{thm:ldp}; the average computing time for Theorem \ref{thm:ldp} for $\varepsilon = 0.5$ is 133s, while for Theorem \ref{thm:lip} this is 0.0206s.  With regards to the utility $\recht{I}(X;Y)$, since $\varepsilon$-LDP implies $\varepsilon$-LIP, the optimal $\varepsilon$-LIP protocol will have better utility than the optimal $\varepsilon$-LDP protocol. However, as can be seen from the figure, the difference in utility is relatively low.

Note that for bigger $\varepsilon$, both the difference in computing time and the difference in $\recht{I}(X;Y)$ between LDP and LIP become less. This is because of the probabilistic relation between $S$ and $X$, for $\varepsilon$ large enough, any sanitisation protocol satisfies $\varepsilon$-LIP and $\varepsilon$-LDP. This means that as $\varepsilon$ grows, the resulting polytopes will have less defining inequalities, hence they will have less vertices. This results in lower computation times, which affects LDP more than LIP. At the same time, the fact that every protocol is both $\varepsilon$-LIP and $\varepsilon$-LDP will result in the same optimal utility.

\begin{figure}
\includegraphics[width = \linewidth]{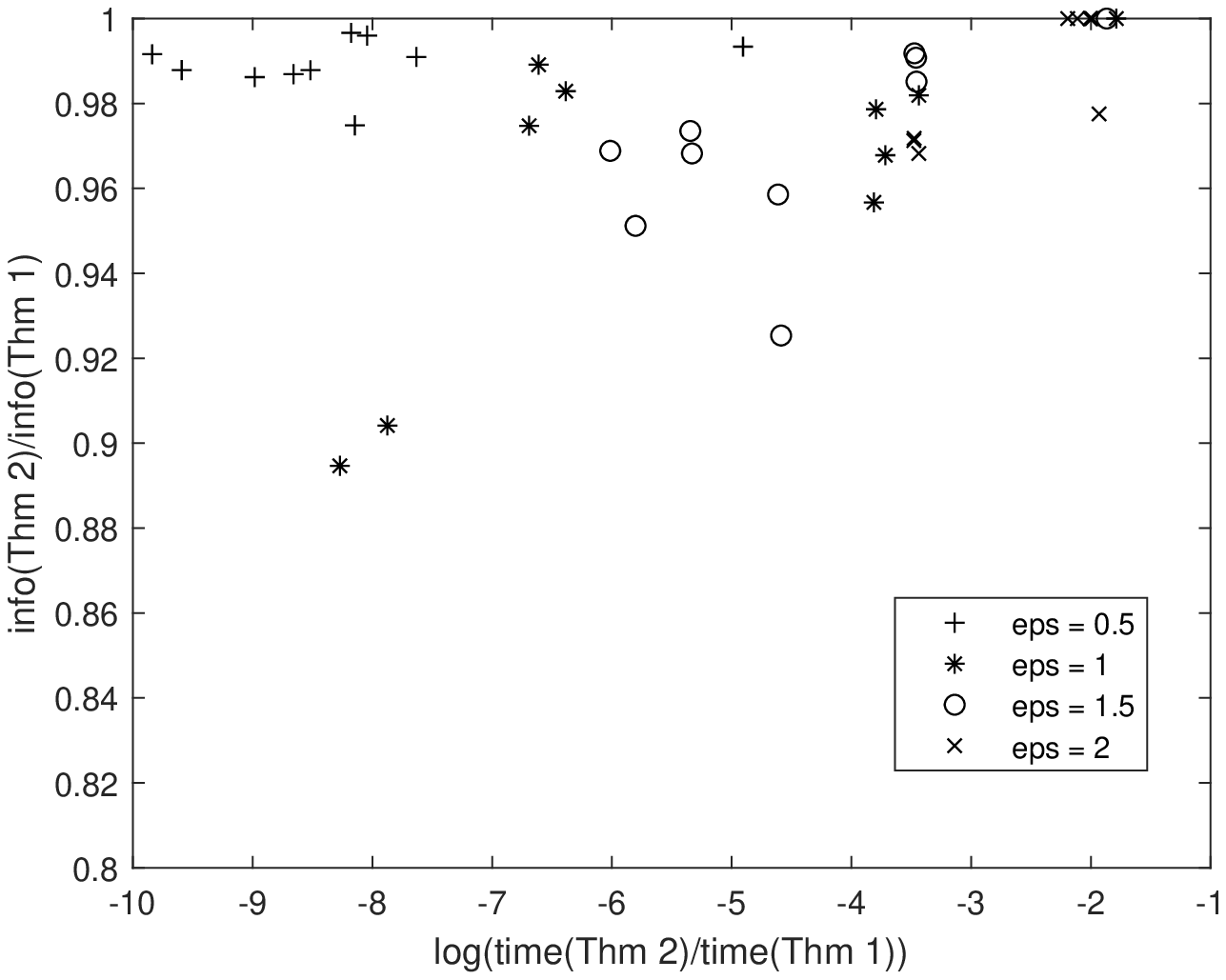}
\centering
\captionsetup{font={footnotesize,rm},justification=centering,labelsep=period}
\caption{Comparision of computation time and $\recht{I}(X;Y)$ for $\varepsilon$-LDP protocols found via Theorem \ref{thm:ldp} and $\frac{\varepsilon}{2}$-LIP protocols found via Theorem \ref{thm:lip}, for random $\recht{p}_{S,X}$ with $c = 2$, $a = 5$, and $\varepsilon \in \{0.5,1,1.5,2\}$.\label{fig:ldplip2}}
\end{figure}

In Figure \ref{fig:ldplip2}, we compare optimal $\frac{\varepsilon}{2}$-LDP protocols to optimal $\varepsilon$-LIP protocols. Again, LIP is significantly faster than LDP. Since $\varepsilon$-LIP implies $\frac{\varepsilon}{2}$-LDP, the optimal $\frac{\varepsilon}{2}$-LDP has higher utility; again the difference is low.

\begin{figure}
\centering
\includegraphics[width = \linewidth]{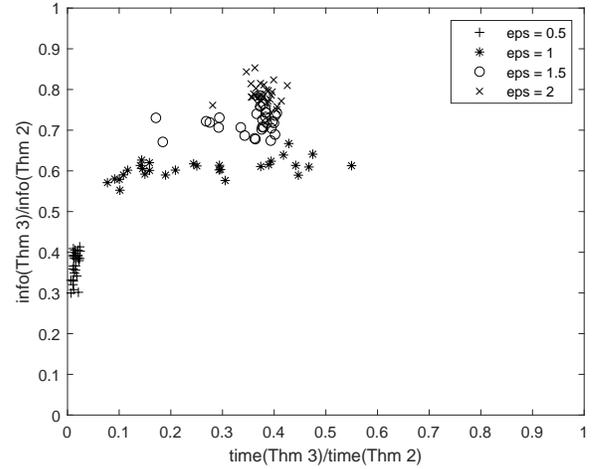}
\captionsetup{font={footnotesize,rm},justification=centering,labelsep=period}
\caption{Comparison of computation time and $\recht{I}(X;Y)$ for $\varepsilon$-(SR)LIP-protocols found via Theorems \ref{thm:lip} and \ref{thm:budget}, for random $\recht{p}_{S,X}$ with $c = 2$, $a_1 = a_2 = 3$, $a_3 = 4$, and $\varepsilon \in \{0.5,1,1.5,2\}$.\label{fig:rslip}}
\end{figure}

We also perform similar comparisons for multiple attributes, for $c = 2$, $a_1 = a_2 = 3$ and $a_3 = 4$, comparing the methods of Theorems \ref{thm:lip} and \ref{thm:budget}. The results are presented in Figure \ref{fig:rslip}. As one can see, Theorem \ref{thm:budget} is significantly slower, with Theorem \ref{thm:lip} being on average $476$ times as fast. There is a sizable difference in utility, caused on one hand by the fact that $\varepsilon$-SRLIP is a stricter privacy requirement than $\varepsilon$-LIP, and on the other hand by the fact that Theorem \ref{thm:budget} does not give us the optimal $\varepsilon$-SRLIP protocol.

\section{Conclusions and future work}

Local data sanitisation protocols have the advantage of being scalable for large numbers of users. Furthermore, the advantage of using differential privacy-like privacy metrics is that they provide worst-case guarantees, ensuring that the privacy of every user is sufficiently protected. For both $\varepsilon$-LDP and $\varepsilon$-LIP we have found methods to find optimal sanitisation protocols.  Within this setting, we have found that $\varepsilon$-LIP has two main advantages over $\varepsilon$-LDP. First, it fits better within the privacy funnel setting, where the distribution $\recht{p}_{S,X}$ is (at least approximately) known to the estimator. Second, finding the optimal protocol is significantly faster than under LDP, especially for small $\varepsilon$. If one nevertheless prefers $\varepsilon$-LDP as a privacy metric, then it is still worthwile to find the optimal $\frac{\varepsilon}{2}$-LIP protocol, as this can be found significantly faster, at a low utility cost.

In the multiple attributes setting, we have shown that $\varepsilon$-SRLIP is a more sensible privacy metric than $\varepsilon$-LIP, since without this requirement a protocol can lose all its privacy protection in the presence of side channels. Unfortunately, however, experiments show that we pay for this both in computation time and in utility. Nevertheless, because of the robustness of $\varepsilon$-SRLIP, it remains the preferred privacy notion in this setting.

For further research, two important avenues remain to be explored. First, the aggregator's knowledge about $\recht{p}_{S,X}$ may not be perfect, because they may learn about $\recht{p}_{S,X}$ through observing $(\vec{S},\vec{X})$. Incorporating this uncertainty leads to robust optimisation \cite{bertsimas2017data} , which would give stronger privacy guarantees. Second, it might be possible to improve the method of obtaining $\varepsilon$-SRLIP protocols via Theorem \ref{thm:budget}. Examining its proof shows that lower values of $\varepsilon^j$ may suffice to still ensure $\varepsilon$-SRLIP. Furthermore, the optimal choice of $(\varepsilon^j)_{j \leq m}$ such that $\sum_j \varepsilon^j = \varepsilon$ might not be $\varepsilon^j = \frac{\varepsilon}{m}$. However, it is computationally prohibitive to perform the vertex enumeration for many different choices of $(\varepsilon^j)_{j \leq m}$, and as such a new theoretical approach is needed to determine the optimal $(\varepsilon^j)_{j \leq m}$ from $\varepsilon$ and $\recht{p}_{S,X}$.

\section*{Acknowledgements}
This work was supported by NWO grant 628.001.026 (Dutch Research Council, the Hague, the Netherlands). The author thanks Jasper Goseling and Boris \v{S}kori\'{c} for helpful discussions.

\appendices

\section{Proof of Theorem \ref{thm:budget}} \label{app:proof}

For $J \subset \{1,\cdots, m\}$ and $j \in \{1,\cdots,m\}$, we write $J[j] := J \cup \{1,\cdots,j-1\}$. Furthermore, we write $\mathcal{X}^{\backslash J} = \prod_{j \notin J} \mathcal{X}^j$, and its elements as $x^{\backslash J}$. We write $\varepsilon := \sum_j \varepsilon^j$. We then have
\begin{align}
\recht{p}_{y|s,x^J} &= \sum_{x^{\backslash J}} \recht{p}_{y|x}\recht{p}_{x^{\backslash J}|s,x^J} \\
&= \recht{p}_{y^J|x^J}\sum_{x^{\backslash j}} \left(\prod_{j \notin J} \recht{p}_{y^j|x^j}\right)\recht{p}_{x^{\backslash J}|s,x^J} \\
&= \recht{p}_{y^J|x^J}\sum_{x^{\backslash j}} \prod_{j \notin J} \recht{p}_{y^j|x^j}\recht{p}_{x^j|s,x^{J[j]}} \\
&= \recht{p}_{y^J|x^J}\prod_{j \notin J} \sum_{x^j} \recht{p}_{y^j|x^j}\recht{p}_{x^j|s,x^{J[j]}} \\
&= \recht{p}_{y^J|x^J}\prod_{j \notin J}\recht{p}_{y^j|s,x^{J[j]}} \\
&\leq \recht{p}_{y^J|x^J}\prod_{j \notin J}\textrm{e}^{\varepsilon^j}\recht{p}_{y^j|x^{J[j]}} \\
&\leq \textrm{e}^{\varepsilon}\recht{p}_{y^J|x^J}\prod_{j \notin J}\recht{p}_{y^j|x^{J[j]}} \\
&= \textrm{e}^{\varepsilon} \recht{p}_{y|x^J}.
\end{align}
The fact that $\textrm{e}^{-\varepsilon} \recht{p}_{y|x^J} \leq \recht{p}_{y|s,x^J}$ is proven analogously.

\begin{thebibliography}{99}
\bibitem{avis1992pivoting} D. Avis and K. Fukuda, ``A pivoting algorithm for convex hulls and vertex enumeration of arrangements and polyhedra,''  \emph{Discrete and Computational Geometry}, vol. 8, no. 1, pp. 174--190, 1992.
\bibitem{barany2000polytopes} I. Bárány and A. Pór, ``On 0-1 Polytopes with Many Facets,'' \emph{Advances in Mathematics}, vol. 161, no. 2, pp. 209--228, 2001.
\bibitem{bertsimas2017data} D. Bertsimas, V. Gupta, and N. Kallus, ``Data-driven robust optimization,'' \emph{Mathematical Programming}, vol. 167, no. 2, pp. 235--292, 2017.
\bibitem{calmon2017principal} F.P. Calmon et al., ``Principal Inertia Components and Applications,'' \emph{IEEE Transactions on Information Theory}, vol. 63 no. 8, pp. 5011--5038, 2017.
\bibitem{cuff2016differential} P. Cuff and L. Yu,  ``Differential Privacy as a Mutual Information Constraint,'' \emph{ACM SIGSAC Conference on Computer and Communications Security 2016}, pp. 43--54, 2016.
\bibitem{ding2019submodularity} N. Ding and P. Sadeghi, ``A Submodularity-based Agglomerative Clustering Algorithm for the Privacy Funnel,'' Unpublished preprint, \emph{arXiv:1901.06629}, 2019 (retrieved: February, 2020).
\bibitem{dwork2006calibrating} C. Dwork, F. McSherry, K. Nissim, and A. Smith, ``Calibrating noise to sensitivity in private data analysis,'' \emph{Theory of Cryptography Conference}, pp. 265--284, 2006.
\bibitem{erlingsson2020encode} Ú. Erlingsson et al., ``Encode, Shuffle, Analyze Privacy Revisited: Formalizations and Empirical Evaluation,'' Unpublished preprint, \emph{arXiv:2001.03618}, 2020 (retrieved: February, 2020).
\bibitem{jiang2019local} B. Jiang, M. Li, and R. Tandon, ``Local Information Privacy with Bounded Prior,'' \emph{IEEE International Conference on Communications}, pp. 1--7, 2019.
\bibitem{kairouz2014extremal} P. Kairouz, S. Oh, and P. Viswanath, ``Extremal Mechanisms for Local Differential Privacy,'' \emph{Advances in Neural Information Processing Systems}, vol. 27, pp. 2879–2887, 2014.
\bibitem{kasiviswanathan2011what} S. P. Kasiviswanathan, H. K. Lee, K. Nissim, S. Raskhodnikova, and A. Smith, ``What can we learn privately?,'' \emph{SIAM Journal of Computing}, vol. 40, no. 3, pp. 793--826,2011.
\bibitem{kifer2014pufferfish} D. Kifer and A. Machanavajjhala, ``Pufferfish: A Framework for Mathematical Privacy Definitions,'' \emph{ACM Transactions on Database Systems}, vol. 39, no 1., pp. 1--36, 2014.
\bibitem{lopuhaazwakenberg2019information} M. Lopuhaä-Zwakenberg, B. \v{S}kori\'{c}, and N. Li, ``Information-theoretic metrics for Local Differential Privacy protocols,'' Unpublished preprint, \emph{arXiv:1910.07826}, 2019 (retrieved: February, 2020).
\bibitem{prasser2014arx} F. Prasser, F. Kohlmayer, R. Lautenschläger, and K. A. Kuhn, ``ARX - A Comprehensive Tool for Anonymizing Biomedical Data,'' \emph{AMIA Annual Symposium Proceedings}, pp. 984--993, 2014.
\bibitem{rassouli2017perfect} B. Rassouli and D. Gündüz, ``On Perfect Privacy and Maximal Correlation,'' Unpublished preprint, \emph{arXiv:1712.08500}, 2017 (retrieved: February, 2020).
\bibitem{salamatian2020privacy} S. Salamatian, F.P. Calmon, N. Fawaz, A. Makhdoumi, and M. Médard, ``Privacy-Utility Tradeoff and Privacy Funnel,'' Unpublished preprint, \url{http://www.mit.edu/~salmansa/files/privacy_TIFS.pdf}, 2020 (retrieved: February, 2020).
\end{thebibliography}
\end{document}